\documentclass{aa}  

\usepackage{graphicx}
\usepackage{txfonts}
\begin{document} 

   \title{Variable emission mechanism of a Type IV radio burst}
        \titlerunning{Variable emission mechanism of a Type IV radio burst}

   \author{D. E. Morosan \inst{1}
          \and
          E. K. J. Kilpua \inst{1} 
          \and
          E. P. Carley \inst{2,3}
          \and
          C. Monstein \inst{4}
          }

   \institute{Department of Physics, University of Helsinki, P.O. Box 64, Helsinki, Finland. \\
              \email{diana.morosan@helsinki.fi}
         \and
             School of Physics, Trinity College Dublin, Dublin 2, Ireland.
           \and
             Astronomy \& Astrophysics Section, School of Cosmic Physics, Dublin Institute for Advanced Studies, 31 Fitzwilliam Place, Dublin, Ireland.
            \and
            Institute for Particle Physics and Astrophysics, ETH Z\"{u}rich, Z\"{u}rich, Switzerland.\\
             }

   \date{Received ; accepted }

 
  \abstract
   {The Sun is an active star and the source of the largest explosions in the solar system, such as flares and coronal mass ejections (CMEs). Flares and CMEs are powerful particle accelerators that can generate radio emission through various emission mechanisms.}
  {CMEs are often accompanied by Type IV radio bursts that are observed as continuum emission in dynamic spectra at decimetric and metric wavelengths, but their emission mechanism can vary from event to event. Here, we aim to determine the emission mechanism of a complex Type IV burst that accompanied the flare and CME on 22 September 2011.}
   {We used radio imaging from the Nan{\c c}ay Radioheliograph, spectroscopic data from the e-Callisto network, ARTEMIS, Ondrejov, and Phoenix3 spectrometers combined with extreme-ultraviolet observations from NASA's Solar Dynamic Observatory to analyse the Type IV radio burst and determine its emission mechanism.}
   {We show that the emission mechanism of the Type IV radio burst changes over time. We identified two components in the Type IV radio burst: an earlier stationary Type IV showing gyro-synchrotron behaviour, and a later moving Type IV burst covering the same frequency band. This second component has a coherent emission mechanism. Fundamental plasma emission and the electron-cyclotron maser emission are further investigated as possible emission mechanisms for the generation of the moving Type IV burst.}
   {Type IV bursts are therefore complex radio bursts, where multiple emission mechanisms can contribute to the generation of the wide-band continuum observed in dynamic spectra. Imaging spectroscopy over a wide frequency band is necessary to determine the emission mechanisms of Type IV bursts that are observed in dynamic spectra.}


\maketitle

%

\section{Introduction}

{The Sun is the source of the most powerful explosions in the solar system, such as flares and coronal mass ejections (CMEs). These phenomena are often associated with accelerated electrons that in turn generate radio emission through various emission mechanisms.}

{CMEs generally erupt when twisted or sheared magnetic fields in the corona become unstable. They are best observed in white-light coronagraphs, such as the Large Angle and Spectrometric Coronagraph \citep[LASCO;][]{br95} on board the Solar and Heliospheric Observatory (SOHO), which records photospheric light scattered by coronal electrons. In white light, CMEs usually consist of a bright core and a cavity surrounded by a bright compression front. The cavity is believed to correspond to an expanding flux rope \citep{ch97, au10} that is an integral part of a CME eruption \citep{vo13}. CMEs can be powerful drivers of plasma shocks that accelerate electrons to high energies. These electrons in turn generate bursts of radiation at metre and decimetre wavelengths through the plasma emission mechanism, such as Type II radio bursts \citep{kl02} and herringbones \citep{ro59,ca13, mo18}. CMEs can also be accompanied by continuum emission at decimetric and metric wavelengths, so-called Type IV radio bursts, which can have either stationary or moving sources and various emission mechanisms \citep[for a review, see][]{ba98}. }

%
   \begin{figure*}[ht]
   \centering
          \includegraphics[width=14cm, trim = 0px 20px 0px 20px]{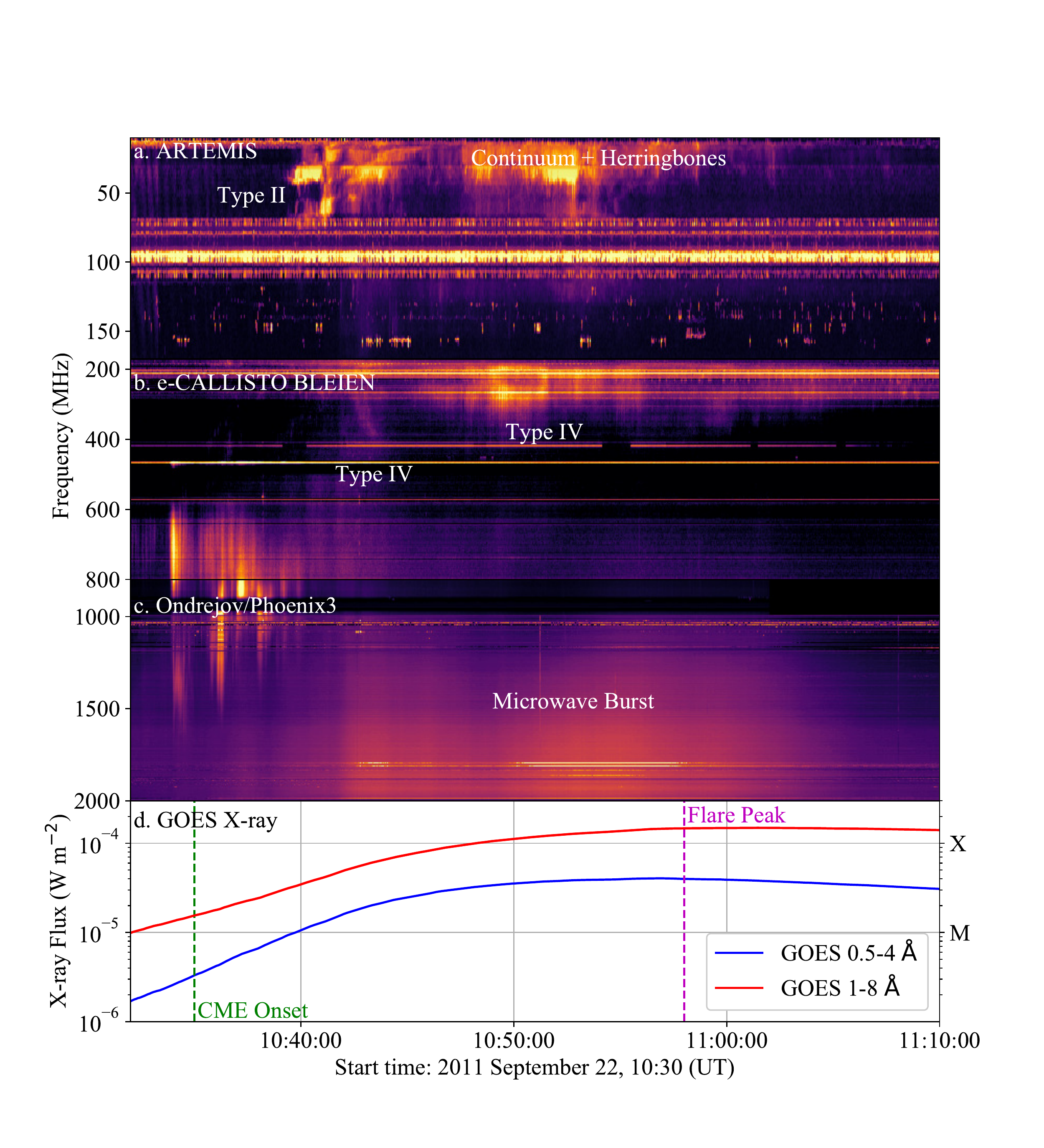}
      \caption{Radio bursts associated with the 22 September 2011 flare and CME. (a) Type II and continuum emission observed by the ARTEMIS spectrograph at frequencies <200~MHz. (b) Type IV bursts observed by the e-CALLISTO BLEIEN dish at frequencies 200--400~MHz and at frequencies (>600~MHz). (c) Microwave burst observed by Ondrejov Spectrograph (0.8--1~GHz) and Phoenix3 spectrometer (1--2~GHz). (d) GOES X-ray light curve of the X1.4 flare on 22 September 2011. The green dashed line denotes the CME onset, which is defined as the time when the CME expands beyond a distance of 1.2~R$_\odot$ , and the purple line denotes the peak time of the flare. }
         \label{fig1}
   \end{figure*}
%

{Of particular interest are moving Type IV radio bursts, which were first classified by \citet{bo57} as broadband radio sources moving outwards from the Sun. First studies suggested that moving Type IV bursts are emitted by synchrotron or gyro-synchrotron emitting electrons that are trapped inside CME loops \citep{bo68,du73}. However, since their discovery, a number of stationary Type IV radio sources have been observed that are believed to be generated by the plasma emission mechanism \citep{we63, be76}, while some moving Type IV bursts have also been identified as plasma emission \citep{ga85}. \citet{ba01} were the first to report the existence of a radio CME, which was observed as an ensemble of loop structures imaged by the Nan{\c c}ay Radioheliograph. The loops extended to distances of up to 3~R$_\odot$ at 164~MHz and were made visible at radio wavelengths by synchrotron emitting electrons.}

{One of the key interests in radio observations of gyro-synchrotron emission and moving Type IV bursts is that they might provide a tool for estimating the magnetic field strengths in CMEs \citep{go87, ba01, ma07, ba14, ca17}. The magnetic field is a key parameter in CME eruption dynamics and its subsequent geo-effectivity, but it has only rarely been measured.  A range of magnetic field strength values has been obtained from moving Type IV observations, which generally vary with height in the corona. For example, Maia et al. (2007) found magnetic field strength values of <1~G at 2~R$_\odot$, while Tun \& Vourlidas (2013) estimated values of 5--15~G at 1.5~R$_\odot$. Moving Type IV bursts do not always accompany CMEs and are considered to be a rare phenomenon because only 5\% of the CMEs were accompanied by moving Type IV bursts based on a spectroscopic study by \citet{ge86}. However, imaging of radio bursts during CMEs has not been considered so far to determine the fraction of moving broadband sources associated with CMEs that may not be obvious in dynamic spectra. It is also not yet clear what the CME characteristics are that are linked to the generation of Type IV bursts, and which emission mechanism is indeed responsible for this emission \citep{li18}. The emission mechanism of Type IV bursts seems to vary on a case-by-case basis.  }

{In this paper, we present observations of the radio emission associated with a CME on 22 September 2011 and show that the accompanying Type IV radio burst shows two components with different emission mechanisms. We also aim to determine possible emission mechanisms that might be responsible for the complex Type IV emission observed. In Section 2 we give an overview of the observations and data analysis techniques. In Section 3 we present the results of the Type IV analysis, which are further discussed in Section 4.  }

%
   \begin{figure*}[ht]
   \centering
          \includegraphics[width=16cm, trim = 30px 50px 0px 20px]{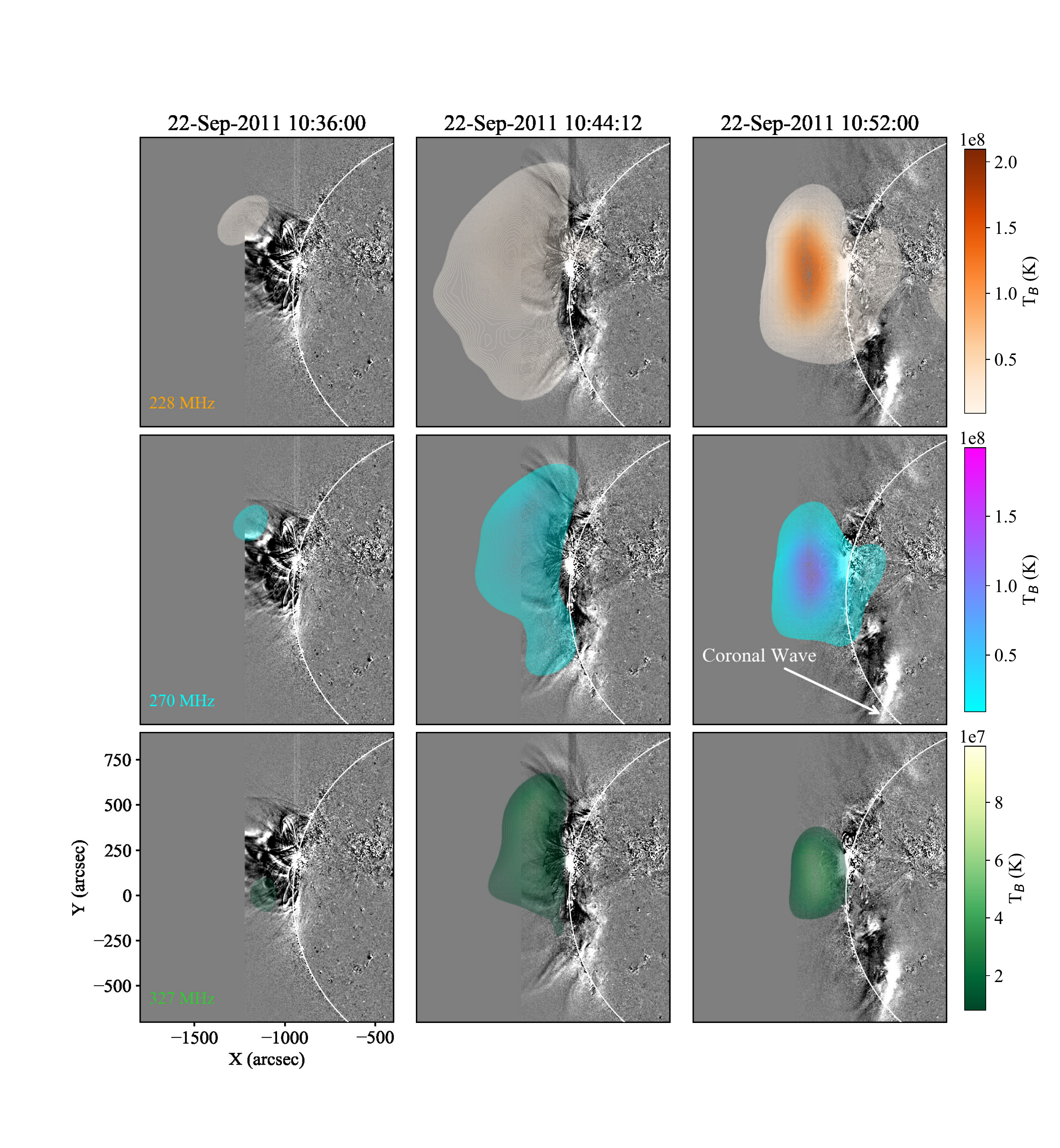}
      \caption{Radio source locations observed by NRH associated with the 22 September 2011 CME. The radio sources (filled contours) are overlaid on AIA 211~\AA~ running-difference images. The radio sources are shown from top to bottom at three different frequencies of 228, 270, and 327~MHz and from left to right at three different times of 10:36, 10:44, and 10:52~UT. The sources in the middle panels are very large. The movie accompanying this paper shows the full evolution in time of these radio sources.}
         \label{fig2}
   \end{figure*}
%

\section{Observations and data analysis}

{On 22 September 2011, an X1.4 class flare occurred on the eastern limb of the visible solar disc, which was accompanied by a fast CME. The flare and CME were associated with radio emission at frequencies ranging from a few hundred kHz to tens of GHz, observed by multiple instruments. The flare and CME were accompanied by Type III radio bursts, a Type II radio burst and continuum-like emission at frequencies <100~MHz, and a broad continuum at higher frequencies (Fig. 1). The Type II radio burst, observed by the ARTEMIS-IV Radiospectrograph in Fig. 1a \citep{ko06}, was associated with a CME-driven shock and is discussed in \citet{zu14}. The continuum emission, following the Type II burst in Fig. 1a, consisted of multiple herringbones and was previously analysed by \citet{ca13}. Herringbone bursts represent electron beams that are accelerated by a shock at the CME flank. The continuum emission at frequencies >200~MHz is a Type IV radio burst. }

%
   \begin{figure*}[ht]
   \centering
          \includegraphics[width=15.5cm, trim = 0px 20px 0px 20px]{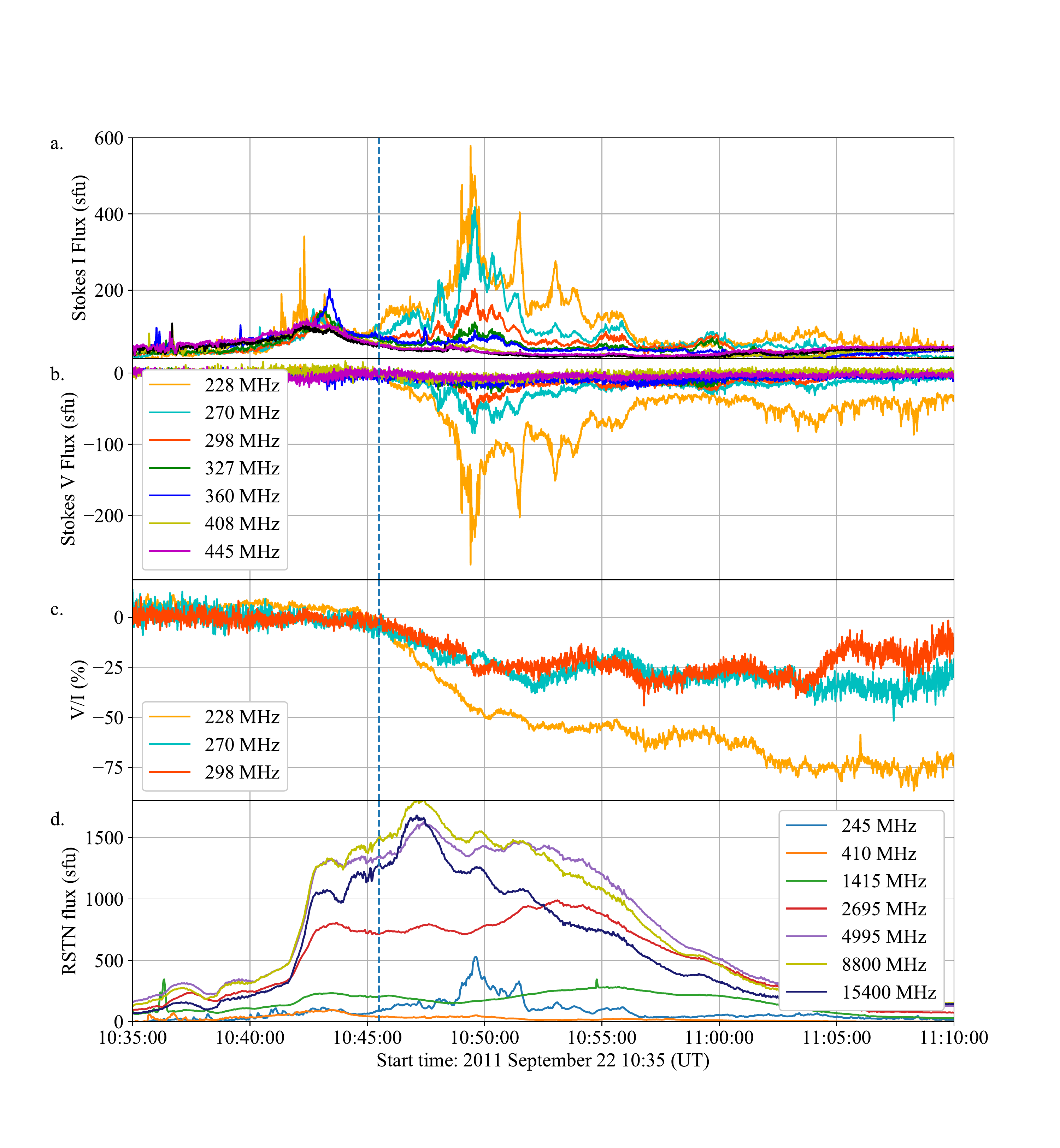}
      \caption{NRH flux densities at individual frequencies from 228 to 445~MHz for Stokes I (a) and Stokes V (b) components. (c) Degree of circular polarisation (V/I) at three NRH frequencies from 228 to 298~MHz. (d) RSTN flux densities at individual frequencies from 245~MHz to 15.4~GHz. The vertical dashed line in each panel separates the two components of the Type IV burst. }
         \label{fig3}
   \end{figure*}


{The Type IV burst accompanying the CME at frequencies between 200--400~MHz in Fig. 1b started at $\sim$10:40~UT and lasted for $\sim$30~minutes. It was observed by some of the solar radio spectrometers that are part of the e-CALLISTO network \citep{be05, be09}, in particular, by the 7~m dish in Bleien, Switzerland (Fig. 1b). Another emission, which appears to be separate from the Type IV emission between 200--400~MHz, was observed at frequencies >600~MHz in Figs. 1b and 1c, starting at $\sim$10:30~UT. The high-frequency part of this emission in Fig. 1c was observed by the Ondrejov RT5 spectrograph, which makes full-disc observations of the Sun in the 0.8--2~GHz range (0.8--1~GHz in Fig. 1c), and by the Phoenix3 spectrometer in the range 1--2~GHz \citep{be09b}. At frequencies >1~GHz, a microwave burst is present that is most likely associated with emission from the active region.}

{In this paper, we focus mainly on the Type IV radio burst at frequencies of 200--400~MHz because this burst was also observed in radio images from the Nan{\c c}ay Radioheliograph \citep[NRH;][]{ke97} at frequencies between 228--445~MHz (Fig. 2). The radio sources in Fig. 2 are represented by filled contours, overlaid on running-difference images in the 211~\AA~passband from the Atmospheric Imaging Assembly \citep[AIA;][]{le12} on board the Solar Dynamics Observatory (SDO).  }

{The NRH flux density was computed to estimate the brightness of the observed Type IV burst and the degree of circular polarization (Fig. 3a-c). The NRH flux was initially estimated over the full solar disc for comparison with calibrated fluxes from the Radio Solar Telescope Network \citep[RSTN;][]{gu79}. We used observations from the Sagamore Hill site of the RSTN network. Sagamore Hill provided us with flux density measurements of the Sun at frequencies of 240, 410, 1415, 2695, 4995, 8800, and 15400~MHz (Fig. 3d). A daily background was subtracted from all RSTN frequencies, so that the flux densities observed are those associated with the flare.}

%
   \begin{figure*}[ht]
   \centering
          \includegraphics[width=15cm, trim = 0px 20px 0px 20px]{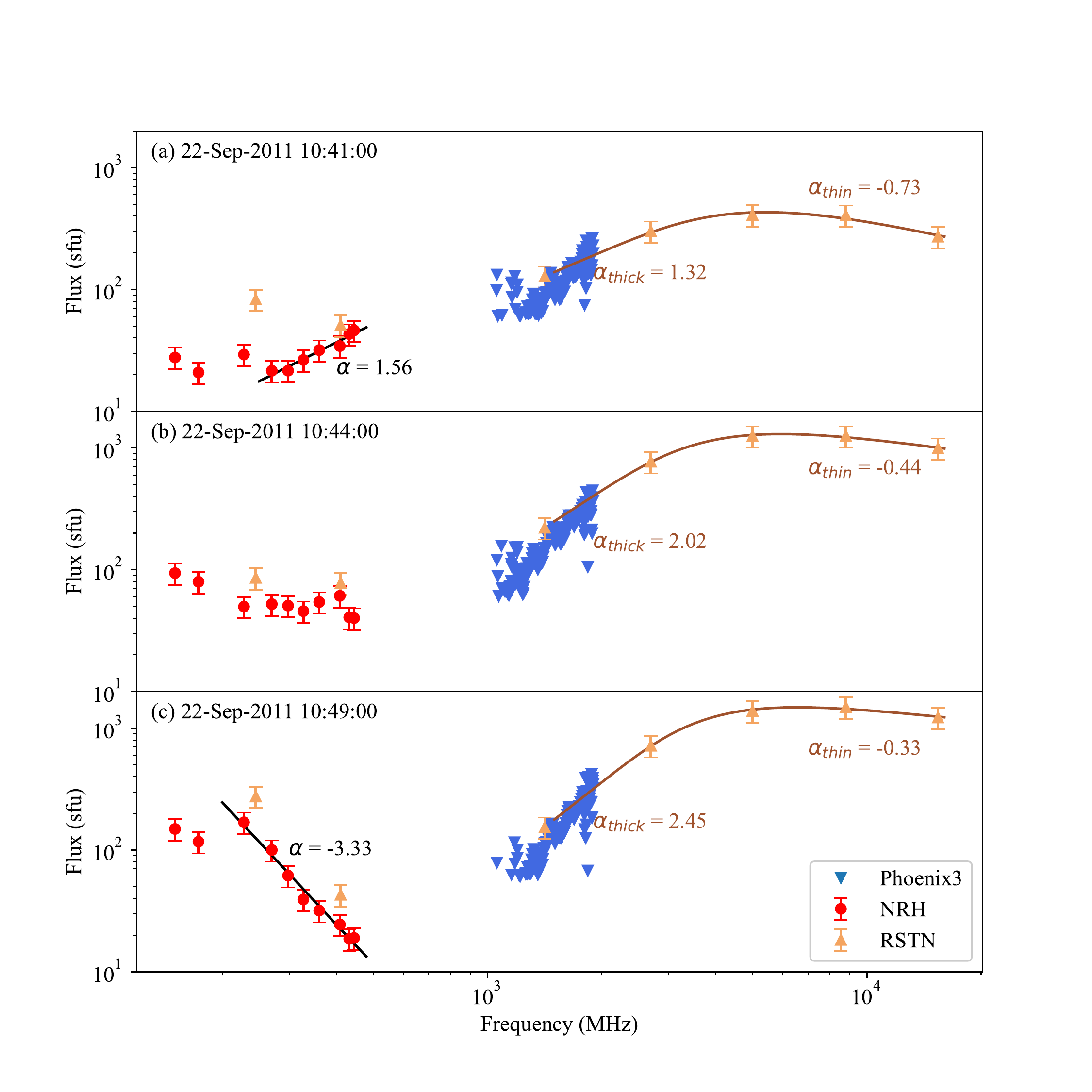}
      \caption{Flux density spectrum during the occurrence of the Type IV radio burst at (a) 10:41~UT, (b) 10:44~UT, and (c)10:49~UT. The NRH flux densities are represented by red circles and are fitted with a power-law function to obtain the spectral index $\alpha$. The RSTN data are represented by orange triangles fitted by a gyrosynchrotron function. The Phoenix3 data are represented by blue inverted triangles.  }
         \label{fig4}
   \end{figure*}
%

\section{Results}

{The Type IV radio burst started at $\sim$10:40~UT, before the flare reached its peak in X-ray emission, and then continued for $\sim$30~minutes until the decay phase of the flare (Fig. 1). Images of the Type IV radio burst show that it initially consisted of a fainter, extended source with brightness temperatures of up to $5\times10^7$~K. This extended source appears to cover the entire lateral extent of the CME at frequencies >228~MHz (middle panels of Fig. 2). The lateral extent of the CME is denoted by the bright and dark features moving northwards and southwards in the AIA 211~\AA~running-difference images. The bright front in the south in Fig. 2 was previously identified as a coronal wave accompanying the CME by \citet{ca13}. The CME has already expanded to the higher corona during the onset of the Type IV burst. Later, the Type IV burst appears to consist of a bright, compact source situated directly on top of the eruption site (right panels of Fig. 2). The brightness temperature of this later source exceeds $10^8$~K at multiple frequencies above 228~MHz. The evolution of the Type IV radio sources in time can also be seen in the movie accompanying this paper.}

{The disc-integrated radio fluxes at NRH frequencies confirm that there are two components: the flux densities are low between 10:40--10:46~UT ($\sim$100~sfu, where 1~sfu = $10^{-22}$~W~m$^2$~Hz$^{-1}$) and less bursty (Fig. 3a). After 10:46~UT, the Type IV emission becomes significantly brighter ($\sim$600~sfu) and consists of bursty components across NRH frequencies between 228--408~MHz (Fig. 3a). The fainter emission between 10:40--10:46~UT is unpolarised, while the brighter, bursty emission after 10:46~UT is up to 80\% circularly polarised (Figs. 3b and 3c). The two components of the Type IV burst described above are split by a vertical dashed line in Fig. 3. The first component might be related to the burst observed at frequencies between 400--870~MHz after 10:40~UT in the dynamic spectrum in Fig. 1b and 1c because the emission appears to continue from the lower (200--400~MHz) to higher (400--870~MHz) frequencies. The second brighter component appears to have no relation to the higher frequency emission in Fig. 1b and 1c. }

{The flux density spectrum of the Type IV burst at NRH frequencies was computed at three different times (red dots in Fig. 4) in order to determine the emission mechanism of the burst. Flux densities were estimated by integrating over the entire emission source observed on the eastern limb in Fig. 2. At higher frequencies (>1~GHz), the flux densities were obtained from the single- frequency RSTN observations at the specified times (orange triangles in Fig. 4). The NRH flux densities show comparable values to the lower frequency RSTN flux densities. At RSTN frequencies, the shape of the spectrum at all three times in Fig. 4 (orange triangles) resembles the shape of a gyro-synchrotron spectrum \citep{st89}. In order to fill the gap between NRH and RSTN frequencies, we used calibrated Phoenix3 data at frequencies 1--2~GHz (for more details about calibration, see \citealt{be09}). The Phoenix3 flux densities at frequencies 1--2~GHz agree with the gyro-synchrotron spectrum of RSTN. }

{At 10:41:00~UT, the flux density increases with frequency in the NRH frequency range (red circles in Fig. 4b), while at 10:49:00~UT, the flux density decreases with frequency in the NRH frequency range >228~MHz (red circles in Fig. 4c). The increase and decrease of flux with frequency has a power-law behaviour. The NRH emission at 150 and 170~MHz was excluded from this analysis because it is associated with the coronal wave in the south that was analysed by \citet{ca13}. The different behaviours at two different times in the NRH flux density spectra also confirm the existence of two separate components in the Type IV emission.}

{The NRH flux density spectrum was fitted with a power-law function, where the spectral index $\alpha$ is defined as the slope of a log-log plot:
\begin{equation}
\alpha = \frac{d \log S}{d \log \nu}
,\end{equation}
where $S$ is the flux and $\nu$ is the observed frequency. }

{At the start, during the faint emission period the spectral index is positive, $\alpha = 1.56$ (Fig. 4a). In terms of gyro-synchrotron spectra, this value is lower than $\alpha_{thick} = 2.9$, which is expected from the optically thick slope from a homogeneous source \citep{du82}, but it falls inside the limits of a recent statistical study by \cite{ni04}, where $\alpha_{thick} = 1.79^{+1.04}_{-0.53}$. It is therefore possible that some of this initial faint emission is gyro-synchrotron in nature and shows a power-law increase at NRH frequencies. At the time of the extended source (middle panels of Fig. 2), the slope of the NRH flux density spectrum is not clearly identifiable (red dots in Fig. 4b), most likely because of various bursts within the extended structure, which is indicative of plasma emission. At this time, it is clear that there is no longer a clear power-law increase in the spectrum, and the spectrum looks rather flat at NRH frequencies (red dots in Fig 4b). Later, at a time where the bright highly polarised source is present above the eruption site (Fig. 4c), the  NRH spectral index is negative, $\alpha = -3.33$.}


%
   \begin{figure*}[ht]
   \centering
          \includegraphics[width=19cm, trim = 20px 20px 0px 20px]{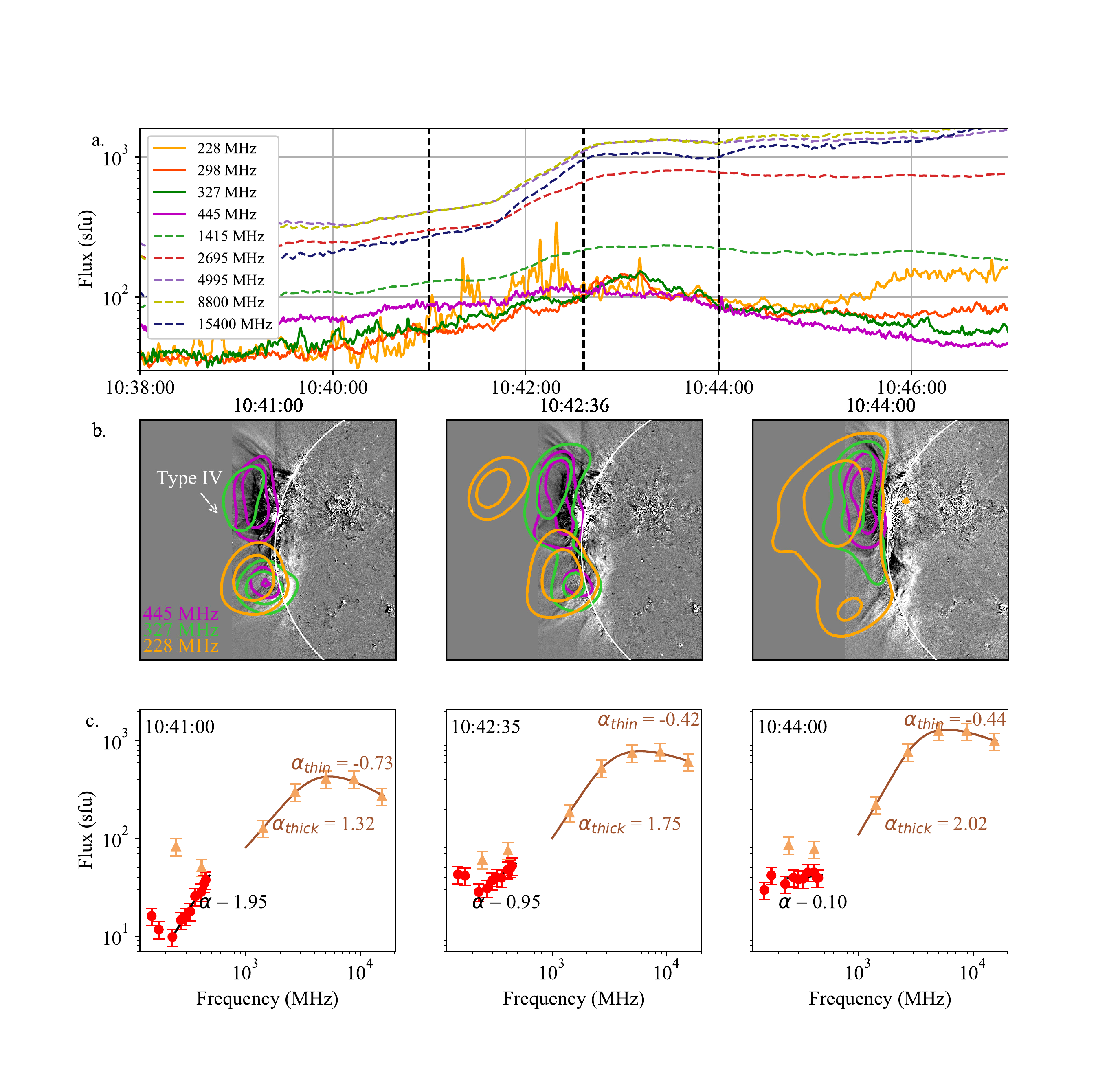}
      \caption{ (a) NRH (solid lines) and RSTN (dashed lines) flux densities at individual frequencies from 228 to 15400~MHz. NRH frequencies show the bump that represents the first component of the Type IV burst. (b) Radio sources are plotted as contours at 30\% and 70\% levels and at three separate frequencies: 228 (orange), 327 (green), and 445~MHz (purple), and at three different times, 10:41:00~UT (left), 10:42:36~UT (middle), and 10:44:00~UT (right), denoted by dashed blue lines in panel (a). The radio contours are overlaid on AIA 211~\AA~running-difference images (c) Flux density spectra at NRH (red dots) and RSTN (blue triangles) frequencies at the same times as in (b). The NRH flux densities were estimated over the northern source denoted Type IV in (b) and it excludes the southern source, which is  associated with the coronal wave that is brighter. NRH fluxes are therefore lower than the RSTN fluxes.   }
         \label{fig5}
   \end{figure*}
%


%
   \begin{figure}[ht]
   \centering
          \includegraphics[width=10cm, trim = 45px 0px 0px 20px]{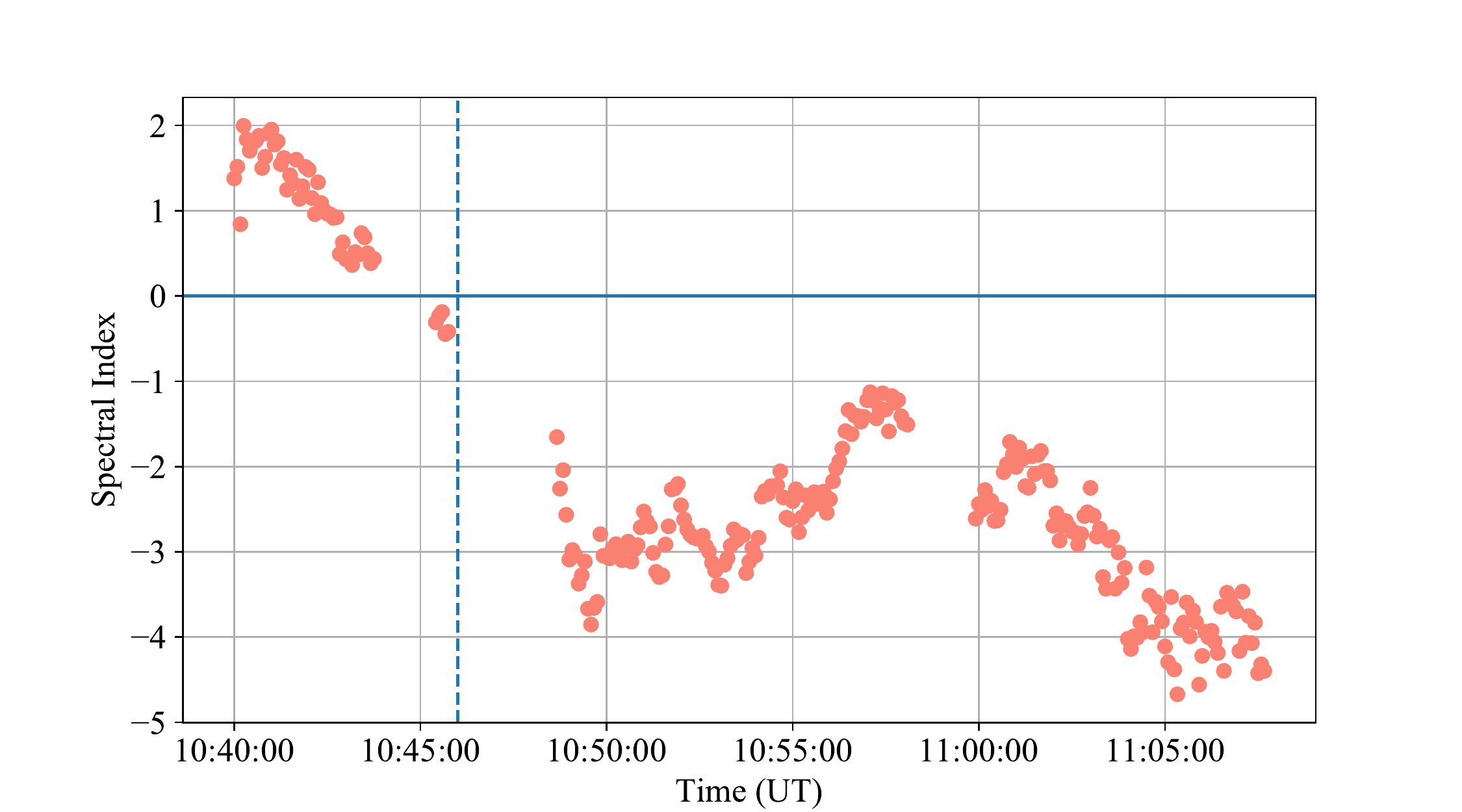}
      \caption{Evolution of the spectral index with time. The vertical dashed line separates the first and second components of the Type IV burst, and the horizontal line denotes the region of flat spectral indices.}
         \label{fig6}
   \end{figure}
%


%
   \begin{figure*}[ht]
   \centering
          \includegraphics[width=19cm, trim = 20px 20px 0px 20px]{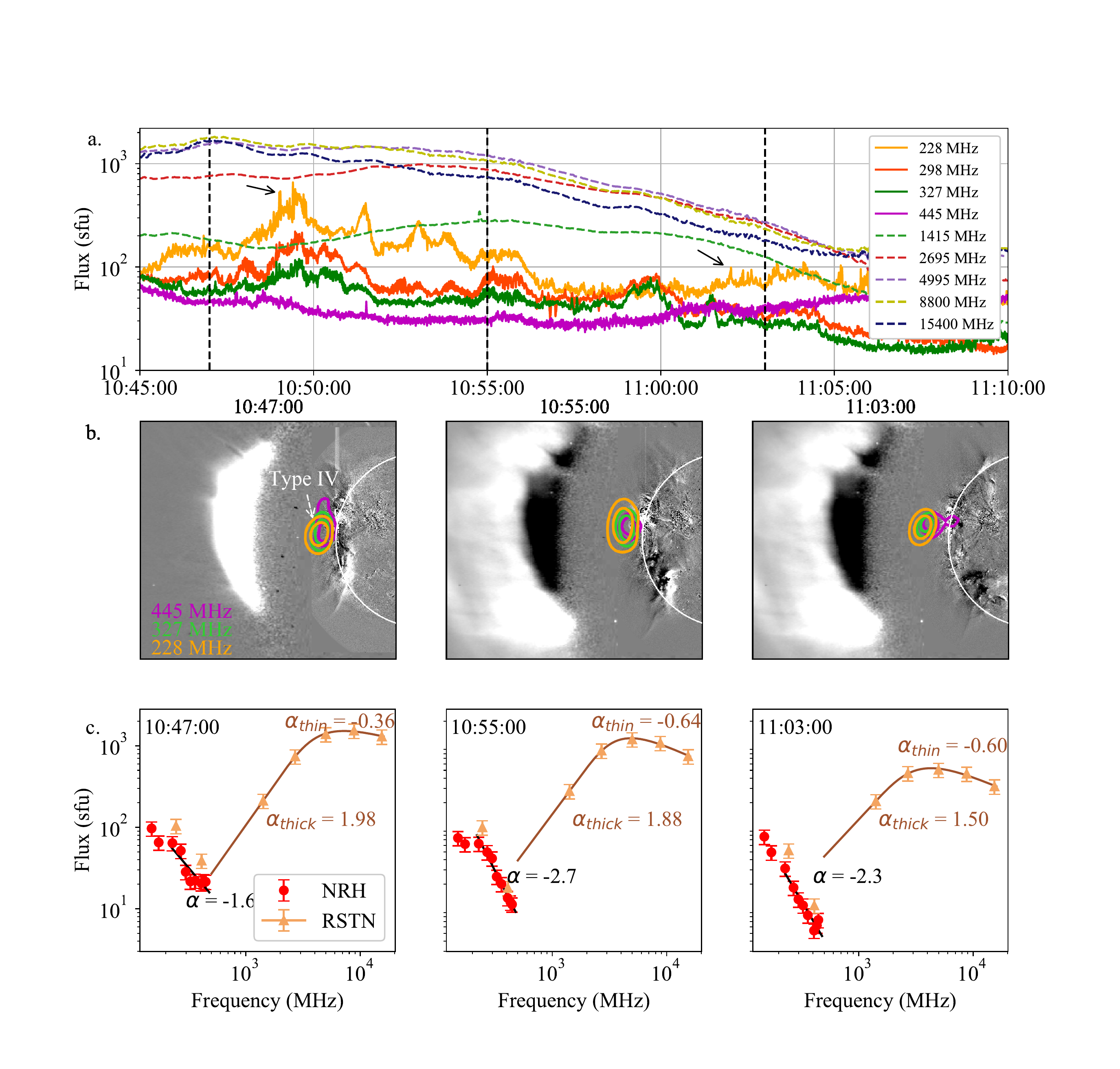}
      \caption{(a) NRH (solid lines) and RSTN (dashed lines) flux densities at individual frequencies from 228 to 15400~MHz. NRH frequencies show the bursty emission that represents the second component of the Type IV burst. (b) Radio sources are plotted as contours at 30\% and 70\% levels and at three separate frequencies: 228 (orange), 327 (green), and 445~MHz (purple), and at three different times, 10:41:00~UT (left), 10:42:36~UT (middle), and 10:44:00~UT (right), denoted by dashed blue lines in panel (a). The radio contours are overlaid on AIA 211~\AA~running-difference images and LASCO C2 running-difference images that show the progression of the CME. (c) Flux density spectra at NRH (red dots) and RSTN (blue triangles frequencies at the same times as in (b). The NRH flux densities were estimated over the northern source denoted Type IV in (b).   }
         \label{fig6}
   \end{figure*}
%

{The RSTN flux density spectrum was fitted with a generic parametric equation for gyro-synchrotron spectra function from \citet{st89} because it has a characteristic shape of a gyro-synchrotron spectrum:
\begin{equation}
S_\nu = S_p \bigg(\frac{\nu}{\nu_p}\bigg)^{\alpha_{thick}} \bigg\{1- \exp \bigg[-\bigg(\frac{\nu}{\nu_p}\bigg)^{\alpha_{thin} - \alpha_{thick}} \bigg] \bigg\}
,\end{equation}
where $S_\nu$ is the flux density, $S_p$ is the peak flux, $\nu_p$ is the peak frequency, $\alpha_{thick}$ is the positive spectral index on the optically thick side of the spectrum, and $\alpha_{thin}$ is the negative spectral index on the optically thin side. The flux density spectrum at RSTN frequencies peaks at a frequency of $\sim$8~GHz, which is close to the median frequency peak of 6.6~GHz found by \cite{ni04}. The optically thick spectral index appears to evolve through time and initially $\alpha_{thick}$ = 1.32 at 10:41~UT. This value is again lower than $\alpha_{thick} = 2.9$, which is expected from the optically thick slope from a homogeneous source \citep{du82}, but it falls inside the limits of the statistical study by \cite{ni04}, where $\alpha_{thick} = 1.79^{+1.04}_{-0.53}$. This value is also close to the positive spectral index, $\alpha = 1.56$, at NRH frequencies. We note that the optically thin spectral index at RSTN frequencies is low, $\alpha_{thin} = -0.73$. However, there are only two data points on the optically thin side of the spectrum, which means that this estimate of the spectral index is less accurate. }

{After 10:44~UT, the emission becomes brighter (>1000~sfu), and the optically thick spectral index becomes larger in magnitude ($\alpha_{thick}$>2), at RSTN frequencies. The optically thick spectral index has a value in the range $\sim$2 for $\sim$15~minutes, after which it decreases in magnitude. Because the RSTN emission peaks at the same frequency at all times in Fig. 4, the emission most likely originates from a microwave burst that occurs in the active region. This is  labelled in Fig. 1. }

{Following the findings presented above, the Type IV burst was therefore split into two components based on its spectral and spatial characteristics, which we further describe below.}


%
   \begin{figure*}[ht]
   \centering
          \includegraphics[width=19cm, trim = 30px 0px 0px 0px]{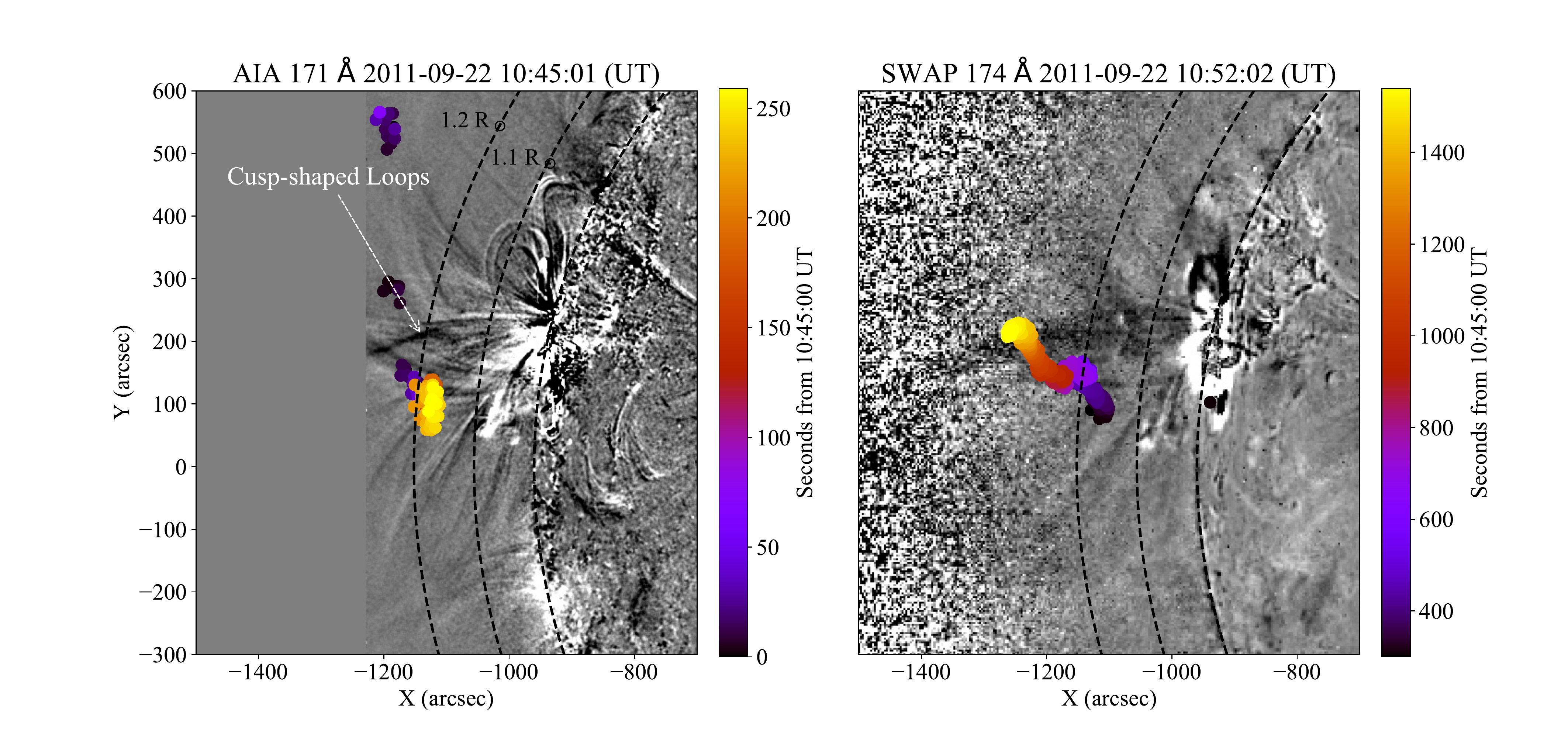}
      \caption{Moving Type IV centroids at 298 MHz split into two phases: a stationary phase (a) and a moving phase (b). The centroids are overlaid on AIA 171~\AA~(a) and SWAP 174~\AA~(b) running-difference images that show the expansion of coronal loops in the negative x-direction behind the CME. The movement of the radio sources in (b) is coincident with the outward growth of the cusp-shaped loop system.} 
         \label{fig7}
   \end{figure*}
%

\subsection{First component of the Type IV burst}

{The first component of the Type IV burst consists of low-intensity unpolarised emission that forms a bump in the flux density curve in Fig. 3 between 10:40--10:46~UT. The fluxes reach a value of ~120~sfu at the peak of the bump, with the exception of a few more prominent peaks at 228 and 327~MHz, which most likely represent individual radio bursts. This bump occurs before the peak time of the flare and while the CME is expanding in the low corona, before the LASCO C2 field of view is reached.}

{The flux density at the time of the first component of the Type IV burst is shown in more detail in a zoom in Fig. 5a. Here, the full-disc single-frequency data from 228~MHz to 15.4~GHz are plotted for the duration of the bump. At the start of this bump, the flux densities at RSTN frequencies also increase (Fig. 5a), but the RSTN fluxes remain high when the NRH emission decreases in flux. We imaged the radio sources at three times during the bump in Fig. 5b. }

{The radio source contours suggest that the extended source on the eastern limb at 10:44~UT consists of two sources: a larger northern source, and a southern source. We note that the southern source that is most prominent at 228~MHz is most likely related to the 150~MHz source studied by \citet{ca13}, which was identified as plasma emission from electrons accelerated by the CME shock. The 228~MHz southern source is co-spatial with the 150~MHz source studied by \citet{ca13} and moves in the same direction as the coronal wave (bright front travelling southwards in Fig. 2 and in the movie accompanying this paper). We therefore excluded this source from the flux density spectrum analysis because this source also contributes to the bright bursty emission at 228 MHz during the bump, which is indicative of plasma emission. The northern source is fainter and less bursty than the southern source.  }

{We therefore identify the northern source as the first component of the Type IV burst we are interested in because it appears consistent in location at all three times (for more details, see the movie accompanying the paper). The large spatial extent of the radio source at 228~MHz at 10:44~UT is most likely a combination of the northern and southern source, and possibly other emissions at higher altitudes. We also note that the 327 and 445~MHz components of the Type IV burst occur at almost the same spatial location (Figs. 2 and 5b). The lack of frequency separation with altitude in observations on the solar limb indicates that this emission was not emitted at the plasma frequency.}

{Flux density spectra during the bump were computed using only the flux density of the northern NRH source (Fig. 5c). Until 10:44~UT, there is a clear power-law increase in the NRH data (red circles). This behaviour is no longer obvious after 10:44~UT, most likely because the second component of the  Type IV burst started, where the flux density increases at lower frequencies in Fig. 5a. This emission most likely peaks above 500~MHz. Above 1000~MHz, the active region microwave burst dominates the dynamic spectrum (Fig. 1c). At the start of the bump (10:41~UT), the spectral index of the northern NRH source is 1.95 (Fig. 5c), which falls inside the limits of the statistical study of gyro-synchrotron indices by \cite{ni04}. }

{The radio sources of the first component of the Type IV burst are stationary, while the CME is expanding outwards. Similar radio sources have been identified before, which remained stationary behind a filament eruption that was interpreted as emission from the current sheet, while simultaneously, moving radio sources were found ahead of the erupting filament, which was interpreted as a moving Type IV burst \citep{vr03}. We first observe stationary gyro-synchrotron sources with locations coincident with post-eruption flare loops (the dark features located below the contours of the Type IV sources in Fig. 5b) and moving radio sources at a later time (the second component of the Type IV burst). In this case, the stationary radio emission most likely originates from gyro-synchrotron electrons that are trapped inside post-eruption flare loops, which remain stationary following the CME expansion. }

{The evolution of the spectral index in time (Fig. 6) shows that the spectral index is initially positive and shows gyro-synchrotron behaviour, after which it gradually decreases to flatness and then it becomes negative following the appearance of the second component of the Type IV burst. The first component of the Type IV burst shows gyro-synchrotron behaviour around 10:41~UT, after which it gradually decreases (Fig. 6). After 10:42~UT, more radio sources appear that contribute to the overall extended source, which is indicative of plasma emission. The gradual change in spectral index towards 0 indicates that the emission mechanism of the Type IV continuum showed changes over time, which has not been reported before. }

\subsection{Second component of the Type IV burst}

{The second component of the Type IV burst consists of high flux densities and bursty emission in Fig. 3, between 10:45--11:10~UT. At 228~MHz, the emission peaks at $\sim$600~sfu, but at higher frequencies (> 408~MHz), the flux density remains below 100~sfu. The second component is separated from the first by the very high degree of circular polarisation of up to 80\% at 228~MHz and up to 40\% at 270 and 298~MHz. The onset of highly polarised emission suggests that the second component corresponds to a different radio source with possibly a different emission mechanism. In addition, the spectral index in Fig. 4c shows a clear power-law decrease, which is opposite to the spectral behaviour of the first component. }

{The flux density at the time of the second component of the Type IV burst is shown in more detail in a zoom in Fig. 7a. Here, the full-disc single-frequency data from 228~MHz to 15.4~GHz are plotted for the duration of the burst. The NRH flux densities appear to have no correlation to the RSTN flux densities in Fig. 5a, indicating that the RSTN microwave burst is not directly related to the NRH sources. Images of radio source contours at three different times show that the radio sources occur above the eruption site, at a time when the CME expanded beyond a distance of 2.2~R$_\odot$ from the solar centre and reached the LASCO field of view (Fig. 7b). The radio sources are plotted as contours at the 30\% and 70\% maximum intensity levels and at three separate frequencies: 228 (orange), 327 (green), and 445~MHz (purple). The radio sources appear co-spatial at all frequencies and show an outwards movement from 10:47 to 11:05~UT.}

{Flux density spectra were computed using the flux densities of the second component of the Type IV burst source at three times: 10:47, 10:55 and 11:03~UT, in Fig. 7c. There is a clear power-law decrease at all times in Fig. 7c, and this trend remains consistent from 10:47~UT until $\sim$11:05~UT (Fig. 6). After 11:05~UT, at 408--445~MHz, the Type IV radio source emission becomes faint and is comparable in brightness to active region and quiet-Sun emissions, as it drops below 10~sfu. At 10:47~UT, we estimate a negative spectral index of -1.6 that also includes contributions from a second source to the north in Fig. 7b. After 10:49~UT (Fig. 4c), the power-law decrease becomes clearer because only the Type IV source is present, and we estimate spectral indices of -3.3 at 10:49~UT and -2.3 at 11:03~UT. }

{The second component of the Type IV burst shows an outward movement through time at all frequencies. The centroid locations of the second component of the Type IV burst are shown in Fig. 8 at 298~MHz. The centroids have been split into two periods: a stationary period at 10:44:20--10:49:20 (Fig. 8a) and a moving period at 10:49:20--11:10:00~UT (Fig. 8b). The centroids are superimposed on an AIA 171~\AA~running-difference image in Fig. 8a and on a 174~\AA~running-difference image from the Sun Watcher using the Active Pixel System detector and Image Processing \citep[SWAP;][]{be06} on board PROBA2 in Fig. 7b. The SWAP image is used to show a larger field of view,  and it has been limb-enhanced to show faint coronal loops beyond the solar limb. }

{The Type IV burst appears stationary for $\sim$5~minutes, but at this time, other sources are present. It is possible that various radio sources are superimposed, and we cannot accurately determine a centroid location. The moving period is characterised by the radio source moving away from the solar disc, and it starts after 10:49~UT, when the spectral index also shows a clear power-law decrease that remains consistent over time (Fig. 7). In both panels, the centroids are coincident with a cusp-shaped loop system that rises to increasing heights with the CME expansion (labelled in Fig. 8). The cusp-shaped loops are located below the CME flux rope, as the CME has already expanded high in the corona (Fig. 7b). It is possible that the Type IV electrons also originate from an acceleration region at the top of these loops, which is located below the flux rope. In this case, the electrons would be accelerated at a reconnecting current sheet that is formed below the flux rope and above the cusp-shaped loops, but evidence of a current sheet is not visible in EUV or white-light images. Larger loops extending to increasing heights (observed in Figure 8) also become visible as reconnection processes occur at successively greater heights. The outward movement of the centroids in Fig. 8b and the growth of the flare loop system is consistent with a reconnecting current sheet that accelerates electrons, where the reconnection sites move higher and higher up along the current sheet \citep[for a review, see][]{fo06}. Radio sources related to a reconnecting current sheet have previously been reported before by \citet{kl00}; \citet{vr03}; \citet{ka05} and \citet{de12}. The radio sources and the growth of the cusp-shaped loop system also have similar outward speeds of 257 and 267~km/s, respectively. }

{The high brightness temperature ($>10^8$ K in Fig. 2) and high degree of circular polarisation (up to 80\% circularly polarised Fig. 3) of the second component of the Type IV burst suggest a coherent emission mechanism, unlike for the first component. \citet{ba14} observed a polarised Type IV burst with similar spectral indices, which they identified as gyro-synchrotron emission in the CME core. However, the Type IV burst studied by \citet{ba14} is located high in the corona at locations coincident with the CME core and is less strongly polarised, as expected for gyro-synchrotron emission with large viewing angles relative to the magnetic field \citep{du82}. The Type IV burst presented here is unlike the burst studied by \citet{ba14}, and it is most likely emitted by a coherent emission mechanism such as plasma emission or the electron cyclotron maser (ECM), which is further discussed in the next section.}

\section{Discussion}

{Following the results outlined in the previous section, we have shown that a Type IV continuum, observed during the flare/CME event on 22 September 2011, has an emission mechanism that changes over time. The Type IV burst consists of at least two components with different characteristics.}

{Based on our analysis, we suggest that the first component of the Type IV burst starts as gyro-synchrotron emission from electrons that are trapped inside post-eruption flare loops. However, the overall emission mechanism changes to plasma emission over time as a result of the appearance of several sources that contribute to the extended structure. The overall source size at 10:44~UT is unusually large, and it is most likely composed of at least three separate radio sources. The spectral index of the first component of the Type IV burst initially shows gyro-synchrotron behaviour and later gradually flattens. A flattening of the optically thick spectral index of gyro-synchrotron emission could be explained by a decrease in the ambient electron density, see \cite{bel96}. Higher-frequency spectral information is necessary to confirm if the emission mechanism of the Type IV burst also changes over a wider frequency range with time. The availability of flux-calibrated spectra or images at higher frequencies (>450~MHz), would also allow for the construction of a complete gyro-synchrotron spectrum and determine if there are gyro-synchrotron emission components inside the CME. }

{We found that the second component of the Type IV burst is a moving Type IV radio burst with consistently negative spectral indices over time, emitted by a coherent emission mechanism. We investigate two scenarios below: plasma emission, and ECM.}


{The high brightness temperature of the moving Type IV burst and the high degree of circular polarisation are indicative of fundamental plasma emission \citep{ro78, me09}. However, the radio burst does not show a spatial separation of radio sources with frequency and it travels upwards in the corona. The location of the centroids of the radio sources is coincident with a set of cusp-shaped coronal loops, most likely located below a reconnecting current sheet at the wake of the erupting CME flux rope. If acceleration does indeed occur at the loop tops and below a current sheet, this emission would be caused by downward-travelling electrons. The outward movement of the centroids is consistent with the reconnection sites moving higher and higher up along the current sheet. This behaviour has been reported by \citet{de12} in the case of radio emission, and has also been observed in EUV images of downflows \citep{sa11}. Fundamental plasma emission could be generated if the electrons were accelerated down the legs of larger and larger loops in the cusp-shaped loop system. However, in this case, we expect a spatial separation of radio sources outwards from the solar limb with decreasing frequency, which is not observed here.}


{Owing to the lack of radio source separation at different frequencies and the unusual spectral indices, it is necessary to investigate the possibility of ECM at the emission site. Properties indicative of ECM emission are the power-law decrease in the flux density spectra and the high degree of circular polarisation after 10:49~UT \citep{wi85, li18}. \citet{li18} observed a stationary Type IV radio burst with similar spectral indices ($\alpha \approx$ -3) consistent through time, and concluded that the most likely emission mechanism to satisfy the observed spectral properties of the Type IV burst is ECM. However, no details were presented regarding the origin of accelerated electrons and the presence of an anisotropic electron pitch-angle distribution for ECM to occur. On the Sun, it is believed that decimetric spikes are a good example of ECM emission, and they appear as clusters of short-duration, narrow-band structures in dynamic spectra \citep[for a review, see][]{fl98}. \citet{fl03} observed spike emission related to the optically thick gyro-synchrotron continuum at GHz frequencies and provided evidence that the trapped electrons producing the gyrosynchrotron emission also have an anisotropic pitch-angle distribution. The flux profiles of the Type IV burst at 228~MHz show bursty features, and some of these features have a full-width at half-maximum (FWHM) duration of <0.5~s, some 20--30~sfu above the background level (denoted by the black arrows in Figure 7a). Some of these features are possibly narrow-band features because they are not observed in the flux profiles at 170 or 270~MHz. Such features are also distinguishable at higher frequencies, but unfortunately, the temporal resolution of the NRH data is 0.25~s. Spikes or finely structured bursts could make up the continuum emission that is observed as the moving Type IV burst. }

{While the moving Type IV burst in this study shows some narrow-band, short-duration structures, these features occur at metric wavelengths and lower frequencies, where plasma emission is believed to be the dominant emission mechanism. Metric spikes have indeed been observed at the starting frequency of Type III bursts, but they are believed to be unrelated to decimetric spikes, and are therefore emitted by the plasma emission mechanism \citep{be82,be96}. However, the short-duration features observed here do not occur simultaneously with Type III radio bursts and are superimposed on the Type IV continuum in the dynamic spectrum. There still remains the question of whether ECM emission is possible at the frequencies and heights of the moving Type IV burst studied here. }

{The main requirement for ECM is that the plasma frequency $\omega_p$ is lower than the electron cyclotron frequency $\Omega_e$ \citep{me91}, where
\begin{equation}
\omega_p = \sqrt{\frac{n_e e^2}{m_e\epsilon_0}}~~ ,
\end{equation}
and
\begin{equation}
\Omega_e= \frac{eB}{m_e}~~ .
\end{equation} 
Here, $n_e$ is the electron plasma density, $B$ is the magnetic field strength, and the remaining quantities are known physical constants. \citet{re15} and \citet{mo16} have shown that this condition can be satisfied in the core of active regions where the magnetic field strength is high. \citet{mo16} showed that ECM is a viable emission mechanism at heights below 1.1~R$_\odot$ and frequencies >500~MHz for a ratio of $\omega_p$/$\Omega_e$=1. The moving Type IV burst presented here starts its outward movement at 1.15~R$_\odot$, higher than the heights predicted by \citet{re15} and \citet{mo16}. \citet{st00} proposed that the ratio $\omega_p$/$\Omega_e$ can be as high as 2 for o- and x-mode emission, in which case it is possible for ECM to occur at greater heights in the corona.}

{For ECM to occur, we also require an anisotropic electron pitch-angle distribution. A way to achieve this is through a loss-cone instability, following magnetically mirrored electrons. However, the sources are located high up in the corona (>1.15~R$_\odot$), below a current sheet or reconnection region, where it is hard to envision how magnetic mirroring would occur. A loss-cone is expected to occur instead above coronal footpoints. }

{A scenario that may explain ECM emission high in the corona was proposed by \citet{tr17}, where ECM can occur during spontaneous magnetic reconnection in narrow current sheets with a guide-field aligned current (i.e. the electric current in the direction of the magnetic field). Simulations show that magnetic reconnection in 2D, with a guide magnetic field perpendicular to the reconnection plane, produces asymmetric electron depletion and electron accumulation regions (exhausts), at each of the outflow regions at the reconnection X-point \citep{le10}. The exhausts have properties relevant to the emission of ECM because they consist of electron depleted regions, which satisfy the condition $\omega_p<\Omega_e$, and with electron momentum distribution functions with steep perpendicular momentum space gradients \citep{tr17}. The steep perpendicular momentum space gradients of the electron distribution function represent the free-energy source for ECM generation. Under strong guide-field conditions, the ECM emission is also predicted to be quite intense. The scenario proposed by \citet{tr17} suits our observations well since the signatures of accelerated electrons occur close to a possible reconnection region (the cusp-shaped loop system). Simulations of current sheets with a guide field in the solar corona have been shown to produce a flux rope \citep{ed17}, and following the flux rope eruption, reconnection at the current sheet might produce ECM emission in the exhaust region, as outlined in \citet{tr17}. }

{The ECM scenario proposed above appears to suit our observations best, but we cannot present any conclusive evidence for either ECM or plasma emission. A statistical study of Type IV bursts accompanying CMEs could help better identify the possible emission mechanisms of Type IV bursts based on their observed properties. Such a study is necessary to determine whether ECM could be a likely emission mechanism for a larger number of Type IV bursts, especially bursts that do not show radio source spatial separation with frequency for limb events. }

\section{Conclusion}

{The flare and CME on 22 September 2011 were accompanied by numerous radio bursts, including a prominent Type IV emission that showed both stationary and moving components. Most interestingly, the emission mechanism of the observed Type IV burst was found to change with time from showing gyro-synchrotron behaviour to plasma or ECM emission, which has not been reported before. }

{Our analysis of the Type IV radio burst highlights the necessity of instruments capable of high-frequency and temporal resolution spectroscopy to complement the imaging capabilities of the NRH and instruments capable of imaging above the NRH frequency range. The Expanded Owens Valley Solar Array (EOVSA) is currently capable of such observations, but it is limited to higher frequency ranges of 1--18~GHz. Other instruments are now available at lower frequencies (<300~MHz), such as the Low Frequency Array (LOFAR) and the Murchinson Widefield Array (MWA), but there is still a need to provide imaging spectroscopy with high resolution at frequencies between 400-1000~MHz. These instruments would provide the possibility of answering outstanding questions on the emission mechanisms of Type IV radio bursts and their potential in estimating CME magnetic fields for space weather purposes. Lastly, ECM emission, if indeed present high in the corona below eruptions, could also be used as an indicator of current sheets in future observations.}

\begin{acknowledgements}{The results presented here have been achieved under the framework of the Finnish Centre of Excellence in Research of Sustainable Space (Academy of Finland grant number 1312390), which we gratefully acknowledge. E.K.J.K. acknowledges the European Research Council (ERC) under the European Union's Horizon 2020 Research and Innovation Programme Project SolMAG 4100103, and Academy of Finland Project 1310445. E.P.C. is supported by the H2020 INFRADEV-1-2017 LOFAR4SW project no. 777442. We would like to acknowledge the NRH, which is funded by the French Ministry of Education and the R\'egion Centre, the ARTEMIS and Ondrejov spectrograph teams, and the e-Callisto and Phoenix3 data access from the Institute of Astronomy, ETH Zurich, and FHNW Windisch, Switzerland. }\end{acknowledgements}

\bibliographystyle{aa} 
\bibliography{Bib_AA.bib} 

\end{document}